\documentclass[twocolumn,showpacs,preprintnumbers,amsmath,amssymb,floatfix]{revtex4}
\usepackage{graphicx}
\usepackage{amsmath}
\usepackage{bm}
\usepackage[usenames,dvipsnames]{color}

\newcommand{\bald}[1]{{\bf #1}}

\newcommand{\eqf}[1]{\begin{equation}\begin{split}#1\end{split}\end{equation}}
\usepackage{dcolumn}

\begin{document}

\title{The Mach cone signal and energy deposition scenarios in linearized hydrodynamics}

\author{R. B. Neufeld}
   \affiliation{Los Alamos National Lab, Theoretical Division, MS B238, Los Alamos, NM 87545, U.S.A.}
\author{Thorsten Renk}
 \affiliation{Department of Physics, P.O. Box 35 FI-40014 University of Jyv\"askyl\"a, Finland}
\affiliation{Helsinki Institute of Physics, P.O. Box 64 FI-00014, University of Helsinki, Finland}

\date{\today}

\begin{abstract}
Particle correlation measurements associated with a hard or semi-hard trigger in heavy-ion collisions may reflect Mach cone shockwaves excited in the bulk medium by partonic energy loss.  This is of great interest because, when compared with theory, such measurements can provide information on the transport properties of the medium.  Specifically, the formation of Mach cone shockwaves is sensitive to the viscosity and speed of sound, as well as the detailed nature of the jet medium interaction.  However, modeling the physics of shockwave excitation to obtain a meaningful comparison with the measured correlations is very challenging since the correlations arise from an interplay of perturbative as well as non-perturbative phenomena at different momentum scales.  In this work we take a step in that direction by presenting a systematic study of the dependence of azimuthal particle correlations on the spatio-temporal structure of energy deposition into the medium.  Our results indicate that detailed modeling of the evolution of an initially produced hard parton and the interaction of this evolving state with the medium is crucial, as both magnitude and shape of the shockwave signal show a strong dependence on the assumptions being made.
\end{abstract}


\maketitle

\section{Introduction}

It is useful to describe the dynamics in ultrarelativistic heavy-ion collisions (URHIC) in terms of the 'bulk' and of 'probes'. The bulk medium, that is, the {\it quark gluon plasma} (QGP), describes QCD matter produced in the collision which is strongly coupled, exhibits collectivity and behaves like a thermalized, near-perfect liquid.  The bulk matter is chiefly responsible for the properties of low transverse momentum ($P_T$) hadron production. On the other hand, there are also hadrons produced at larger momenta which are clearly not thermalized. Such hadrons originate from hard partonic processes which probe such small time and distance scales that they are essentially unmodified by the medium. However, these high $p_T$ partons undergo a final state interaction while they propagate through the medium before hadronization. This 'jet quenching' \cite{Jet1,Jet2,Jet3,Jet4,Jet5,Jet6} has been expected by theory and is experimentally confirmed in measurements of the nuclear suppression factor $R_{AA}$ \cite{PHENIX_R_AA}.

If there is energy loss from a high $p_T$ parton, energy conservation requires that this lost energy flows somewhere. Measurements of particles associated with a high $P_T$ trigger hadron \cite{PHENIX_2pc,STAR_2pc} have given a hint to answer this question: Instead of a back-to-back jet-like correlation structure on the near (trigger) side and the away side as observed in d-Au collisions, the observed correlation shows a surprising splitting of the away side peak into a double-hump structure. This has been early on interpreted as the reaction of the bulk medium to the hard probe in terms of a shockwave \cite{solana} and early phenomenological investigations with a fluid-dynamics inspired approach to propagate the shockwave in the background of an evolving bulk medium and with a full modeling of the trigger bias \cite{Cones1,Cones2,Cones3} have confirmed that a shockwave signature is not erased by the medium-flow induced distortion or by the averaging over many different triggered events, but can indeed account for the correlation observed in the data.  Since then, the focus has been on a more rigorous theoretical formulation of coupling a source of energy and momentum into the hydrodynamical equations using Hard Thermal Loop or AdS/CFT methods \cite{Neufeld:2008fi,Neufeld:2008hs,Neufeld:2008dx,Betz:2008wy,Noronha:2008tg}.

In bringing such proof-of-concept calculations closer to a comparison with data, one of the key questions is the time dependence of the energy and momentum deposition into the medium.  In this paper, we make the assumption that the energy and momentum absorbed by the medium are related by an on-shell condition so that $\bald{u} dE/dt = d\bald{p}/dt$ (see Section \ref{source_section}), where $dE/dt$ and $d\bald{p}/dt$ are the energy and momentum deposition rates, respectively, and $\bald{u}$ is the velocity of the high $p_T$ parton.  For this reason, we focus on the time dependence of the energy deposition, which is potentially driven by many different effects:   First, the strength of the interaction between hard parton and medium depends on the medium density, and in a real heavy ion collision this density varies as a function of space roughly as given by the nuclear overlap and, due to the expansion of the medium, also drops as a function of time. The expansion dynamics therefore tends to lead to less interaction with the medium and hence less energy loss at late times. On the other hand, radiative energy loss in a constant medium of length $L$ has a characteristic $L^2$ dependence due to LPM suppression of near collinear gluon radiation, and this effect tends to increase energy loss at late times, to some degree even in an expanding medium. In addition, in \cite{Neufeld:2009za} it was suggested that gluons radiated from a hard parton subsequently themselves interact with the medium and hence contribute to the energy flow into the medium, leading to a 'crescendo' in the shockwave excitation and large energy deposition at late times. On the other hand, low energy partons undergoing strong energy loss cannot act for a long time as sources of energy, but become absorbed by the medium after a short time already. This effect again tends to lead to small energy deposition at late times, as energy deposited into the medium early on is not available later.

The arguments given so far assume that the source of energy and momentum entering the medium is a single on-shell parton which subsequently undergoes interactions which lead to induced gluon radiation.  However, in a typical hard event, partons are produced with large initial virtualities and even in vacuum evolve into a parton shower where the individual quanta have lower virtualities, and the timescale of the shower evolution is such that it at least partially takes place before a medium is produced. This implies that it may be wrong to think of a single parton initially depositing energy into the medium --- the dynamics may rather be that a developed parton shower acts as a strong source initially, but energy deposition decreases soon as the energy of subleading shower partons is quickly depleted.

It follows from the above that modeling the time-dependence of energy deposition correctly is not a simple and straightforward issue. It is the purpose of this paper to demonstrate that the question is nevertheless highly relevant: Modeling the time-dependence of energy deposition in a different way alters both the magnitude and the shape of the shockwave signal in a significant way. We illustrate this point using different model assumptions for the energy deposition within a constant medium and solving linearized hydrodynamical equations. The paper is organized as follows:  In section \ref{linear_hydro_section} we review the underlying formalism of linearized hydrodynamics, while in section \ref{azimuthal_section} we discuss how to obtain the azimuthal hadron spectrum from the linearized equations of motion.  In section \ref{source_section} we discuss the motivation for the form of the hydrodynamic source term used in our results and show how it depends sensitively on the time-dependence of the energy deposition.  In section \ref{energy_scenarios} we present the different energy deposition scenarios used in our results and discuss the underlying physics assumptions of each one.  The reader only interested in the results of our calculations can skip directly to sections \ref{input_qual}  and \ref{num_results} where specific numerical inputs and resulting azimuthal spectrums are presented.

\section{The Medium Excitation}\label{medium_excitation}

\subsection{Linearized Hydrodynamics}\label{linear_hydro_section}

In what follows we consider a hard parton (parton here refers to the parent parton and the associated secondaries) propagating in an infinite and static QGP.  Additionally, we ignore any net baryon density, which is a reasonable assumption for RHIC energies at mid-rapidity \cite{Bearden:2003fw}.  This parton acts as a source of energy and momentum which is coupled to the linearized hydrodynamic equations of the underlying medium.  The linearized approximation is valid when the energy and momentum density generated by the hard parton is small compared to the equilibrium energy density of the medium.  More will be said on the linearized approximation in Section \ref{results}.

The assumption of an infinite and static QGP is clearly unrealistic for heavy-ion collisions.  However, our purpose here is not to present a study which is directly comparable to experimental data, but rather to show the effect of the time-dependence of energy deposition on the azimuthal particle spectrum associated with a hard parton.  The linearized approximation in a static medium is a good toy model for such a study because the effect of changing parameters such as viscosity or energy deposition scenarios can be easily extracted.  Even within the linearized approximation one could go beyond the static medium, for instance by assuming an underlying Bjorken expansion.  This, however, would introduce complications associated with underlying flow fields and boundary conditions, and is beyond the scope of the current paper.

In the linearized approximation, the effect of the source is to create a local perturbation in the medium, so that the energy-momentum tensor has the linearized form
\eqf{
T^{\mu\nu} = T_0^{\mu\nu} + \delta T^{\mu\nu}
}
where $\delta T^{\mu\nu}$ is the perturbation generated by the source, and $ T_0^{\mu\nu}$ is the equilibrium energy-momentum tensor of the underlying medium.  The fast parton's ability to perturb the medium is encoded in the source term, $J^\nu$ (to be specified below), which couples to the gradient of the energy-momentum tensor
\begin{equation}\label{lin_source}
\partial_\mu \delta T^{\mu\nu} = J^\nu ,
\end{equation}
where $\partial_\mu T_0^{\mu \nu} = 0$.

The equations of motion for a medium coupled to a source in linearized hydrodynamics are discussed in several places (for instance, \cite{solana, Neufeld:2008dx}).  The solution for $\delta T^{\mu\nu}$ in terms of $J^\nu$ is most easily expressed in momentum space by taking the Fourier transform of (\ref{lin_source}).   The result to first order in shear viscosity, $\eta$, for the perturbed energy density, $\delta T^{00} \equiv \delta \epsilon$, and momentum density, $\delta T^{0i} \equiv \bald{g}$, are given by
\begin{eqnarray}\label{eps}
\delta\epsilon ({\mathbf k},\omega) &=& \frac{i k J_L({\mathbf k},\omega)  + J^0({\mathbf k},\omega)(i \omega -  \Gamma_s k^2)}{\omega^2 -  c_s^2 k^2 + i \Gamma_s \omega k^2}, \\
\label{gl}
{\mathbf g}_L ({\mathbf k},\omega) &=& \hat{\mathbf k} g_L = \frac{ i \omega \hat{\mathbf k} J_L({\mathbf k},\omega)+ i c_s^2 {\mathbf k} J^0({\mathbf k},\omega)}{\omega^2 -  c_s^2 k^2 + i \Gamma_s \omega k^2}, \\
\label{gt}
\bald{g}_T({\mathbf k},\omega) &=& \bald{g} - \bald{g}_L = \frac{i{\mathbf J}_T({\mathbf k},\omega)}{\omega +  \frac{3}{4}i \Gamma_s k^2}.
\end{eqnarray}
In the above result, $c_s$ denotes the speed of sound, $\Gamma_s \equiv \frac{4 \eta }{3(\epsilon_0 + p_0)} = \frac{4 \eta }{3 s T}$ is the sound attenuation length, $\epsilon_0$ and $p_0$ are the unperturbed energy density and pressure, respectively and $s$ is the entropy density.  Also, the source and perturbed momentum density vectors are divided into transverse and longitudinal parts: ${\mathbf g} = \hat{\mathbf k} g_L + {\mathbf g}_T$ and ${\mathbf J} = \hat{\mathbf k} J_L + {\mathbf J}_T$, with $\hat{\mathbf k}$ denoting the unit vector in the direction of ${\mathbf k}$.  The position space result for equations (\ref{eps} - \ref{gt}) are obtained by reverse Fourier transform using the general rule
\begin{equation}\label{generalrule}
F({\bf x},t) = \frac{1}{(2 \pi)^4}\int d^3 k \int d \omega \,e^{i \bald{k}\cdot\bald{x} - i \omega t} F({\bf k},\omega).
\end{equation}

\subsection{The Azimuthal Spectrum}\label{azimuthal_section}

Once the source term has been specified, and a solution for $\delta \epsilon$ and $\bald{g}$ is obtained from equations (\ref{eps} - \ref{generalrule}), we will be interested in determining the azimuthal particle spectrum generated by the source.  Ignoring viscous corrections, the induced medium flow velocity is given by
\begin{equation}
\delta\bald{u}(x) = \frac{\bald{g}}{\epsilon_0 + p_0} = \frac{\bald{g}}{\epsilon_0(1 + c_s^2)}
\end{equation}
where we have used $c_s^2  = \partial p/\partial \epsilon \approx  p_0/\epsilon_0$ for $T \gg T_c$ \cite{Heinz:2009xj}.  The total medium four-velocity, which is a sum of the underlying medium and induced velocities, is given by
\eqf{\label{pertu}
(u_0 + \delta u(x))^\mu = \left(1, \frac{\bald{g}}{\epsilon_0(1 + c_s^2)} \right)
}
where $\delta u^0 = 0$ in the limit of a static background.  An expression for $\delta T$ can similarly be found from dimensional considerations.  We write the medium energy density as $\epsilon = A T^4$, where A is some constant, from which one has
\begin{equation}
\epsilon_0 + \delta \epsilon \approx A T_0^4(1  + 4 \frac{\delta T}{T_0})
\end{equation}
leading to
\begin{equation}\label{delT}
\delta T(x) = \frac{\delta \epsilon}{4 \epsilon_0} T_0.
\end{equation}

Having expressions for the flow velocity and temperature, it is now possible to construct the medium's distribution function, which in the Boltzmann limit is given by
\begin{equation}\label{mod_dist}
f(x,p) = e^{-\beta u^\mu p_\mu} = \exp{\left[-\frac{(u_0 + \delta u(x))^\mu p_\mu}{T_0 + \delta T(x)}\right]}.
\end{equation}
Here, $\beta \equiv 1/T$ is the inverse temperature. The distribution is converted into an azimuthal particle distribution by using a Cooper-Frye freeze-out scenario \cite{Cooper:1974mv}.  Consistent with the approach discussed in \cite{Betz:2008wy,Noronha:2008tg}, the final azimuthal particle spectrum for massless particles at mid-rapidity $(y = 0)$ is given by
\begin{equation}\label{cfform}
\frac{d N}{dy \, d\phi}(y = 0) = \int_{p^i_T}^{p^f_T} \frac{d p_T \, p_T}{(2\pi)^3} \int d \Sigma_\mu p^\mu (f(x,p) - f_{0})
\end{equation}
where $\Sigma_\mu$ is the freeze-out hypersurface and
\begin{equation}\label{pT4vec}
p^\mu = (p_T, 0, p_T\sin(\phi),p_T \cos(\phi)).
\end{equation}
The isotropic background contribution, $f_0 = e^{-\beta_0 u_0^\mu p_\mu}$, is subtracted in (\ref{cfform}).  In what follows, it is understood that the source parton propagates along the $\hat{z}$ axis, which also determines the direction of $\phi = 0$ in (\ref{pT4vec}).  We will consider an isochronous freeze-out, as appropriate for a constant medium, in which case $d \Sigma_\mu = d V(1,0)$.

\subsection{The Source Term}\label{source_section}

Nothing has been said up to this point about the form of $J^\nu$ to be used in (\ref{eps} - \ref{gt}).  A common choice is the simple form
\eqf{\label{simpleJ}
J^\nu(x) = \frac{d E}{d t}\delta(\bald{x} - \bald{u} t)U^\nu
}
where $U^\nu \equiv (1,\bald{u})$, $\bald{u}$ is the velocity of the source parton, which is assumed to be at the origin at $t = 0$, and $dE/dt$ is the time-dependent rate of energy loss into the medium (in principle this can be less than the total rate of energy loss, if energy lost from a leading parton is for example redistributed to non-thermalized degrees of freedom). This expression contains an eikonal assumption as it describes a point source propagating with the speed of light on a straight line. The advantage of the source term in (\ref{simpleJ}) is that it trivially conserves energy and momentum at each step in time.  One can see this by integrating both sides of (\ref{lin_source}) over all space, from which it is found for the $\nu = 0$ component
\eqf{
\frac{d}{d t}\int d\bald{x} \, \delta \epsilon = \frac{d E}{d t},
}
which shows that the energy going into the medium is properly accounted for.  A similar exercise holds for the other components of $\nu$ as well.

However, one can go beyond the simple form of (\ref{simpleJ}) and still conserve energy and momentum at each step in time.  Consider adding a total derivative to (\ref{simpleJ})
\eqf{\label{complexJ}
J^\nu(x) = \frac{d E}{d t} U^\nu \delta(\bald{x} - \bald{u} t) \rightarrow \frac{d E}{d t}(U^\nu - \lambda \partial^\nu)\delta(\bald{x} - \bald{u} t)
}
where $\lambda$ is a coefficient with dimension of length.  Again integrating both sides of (\ref{lin_source}) over all space shows that energy and momentum are still properly accounted for.  $\lambda$ acts as a local medium excitation parameter, that is, its contribution integrates to zero globally.  The replacement in (\ref{complexJ}) is motivated by the form of the kinetic theory derived source term for a parton in a perturbative QGP obtained in \cite{Neufeld:2008hs}, and is similar to the source term derived for a quark in a strongly-coupled supersymmetric Yang-Mills plasma \cite{Chesler:2007sv}.  The relativistic limit ($\gamma \gg 1$) of the source derived in \cite{Neufeld:2008hs} can be put into the form
\eqf{\label{ktrel}
J^\nu = \frac{\alpha_s C_2 m_{\rm D}^2}{8 \pi}\left((1,\bald{u})\frac{\gamma }{(\rho^2 + \gamma^2 z_-^2)^{3/2}} - \partial^\nu  \frac{1}{2(\rho^2 + \gamma^2 z_-^2)}\right)
}
where $C_2$ is the Casimir of the source parton, $m_D$ is the Debye screening mass in the medium, $\rho = \sqrt{x^2 + y^2}$ and $z_- = z - u t$ for a source parton propagating in the positive $\hat{z}$ direction and $\gamma = 1/\sqrt{1-u^2}$.

If one considers the source to be localized, (\ref{ktrel}) can be put into the mold of (\ref{complexJ}) by integrating the distributions over all space, and normalizing to a $\delta$ function.  It is helpful to add a damping factor, $e^{-\rho \,m_{\rm D}}$, to the distributions in (\ref{ktrel}) which regulates an infrared divergence that arises when integrating over all space. The form of our damping factor is motivated by the fact that medium induced screening of the hard parton's color fields occurs on the Debye scale.  The damping factor simulates the Debye screening of a Lorentz contracted distribution, hence the form  $e^{-\rho \,m_{\rm D}}$.

We find 
\eqf{
\frac{e^{-\rho \,m_{\rm D}} \gamma }{(\rho^2 + \gamma^2 z_-^2)^{3/2}} \approx 4\pi \, G_0\left(\frac{m_{\rm D}}{2\sqrt{E_p T}}\right)\delta(\bald{x} - \bald{u} t)
}
where $(2 \sqrt{E_p T})^{-1}$ has been introduced as a short distance cutoff ($E_p$ is the energy of the hard parton and $T$ the medium temperature), and $G_0$ is a representation of the incomplete Gamma function
\begin{equation}\label{incgamma}
G_0(z) = \int_{z}^{\infty} dt \frac{e^{-t}}{t}.
\end{equation}
One can also show
\eqf{
 \frac{1}{2(\rho^2 + \gamma^2 z_-^2)} \approx \frac{\pi^2}{\gamma \, m_{\rm D}} \delta(\bald{x} - \bald{u} t).
}
With these approximations, (\ref{ktrel}) is thus written in the form of (\ref{complexJ}) as
\eqf{\label{PcomplexJ}
J^\nu =  \frac{d E}{d t}\left(U^\nu - \lambda \partial^\nu \right)\delta(\bald{x} - \bald{u} t)
}
where
\eqf{\label{lamform}
\lambda =  \frac{\pi}{4 \, \gamma \, m_{\rm D} \,G_0\left(\frac{m_{\rm D}}{2\sqrt{E_p T}}\right)}
}
and
\eqf{
\frac{d E}{d t} = \frac{\alpha_s C_2 m_D^2}{2} G_0\left(\frac{m_{\rm D}}{2\sqrt{E_p T}}\right).
}

Localizing the source term to a $\delta$ function, such as what has been done in (\ref{PcomplexJ}), has the advantage of making the linearized hydrodynamics easier to solve, but also has a physics justification.  Hydrodynamics is a long distance theory which assumes local thermal equilibrium, whereas the energy deposition occurs on short distance scales and is a highly dissipative process.  There is a natural separation of distance scales between the energy deposition and the medium response as described by hydrodynamics (in perturbation theory these scales are $1/(g T)$ and $1/(g^4 T)$, respectively, where $g \ll 1$).  Localizing the source term to a $\delta$ function is consistent with this separation of scales.

As will be seen in the results below, the coefficient $\lambda$ is especially important for generating Mach-like signals in the final azimuthal particle spectrum.  This can be traced back to equations (\ref{eps} - \ref{gt}).  Equation (\ref{gt}) is a diffusion equation and the quantity ${\mathbf g}_T$ is diffusive momentum density generated by the fast parton.  Physically, the diffusive momentum contribution is a wake which flows in the direction of the fast parton's propagation and is not a sound wave.  Previous studies \cite{Betz:2008wy,Betz:2008ka} have shown that the diffusive momentum tends to fill up any double-peak structure in the final spectrum.  On the other hand, equations (\ref{eps}) and (\ref{gl}) describe damped sound waves propagating at speed $c_s$: it's clear that $\delta\epsilon$ and ${\mathbf g}_L$ are the energy and momentum density carried by sound generated by the fast parton and will be responsible for Mach-like signals in the azimuthal spectrum.  When writing a source of the form (\ref{complexJ}) in momentum space we find
\begin{equation}\label{momcomJ}
\begin{split}
J^\nu(\bald{k}) &= \int d^4 x \,e^{-i \bald{k}\cdot\bald{x} + i \omega t} \frac{d E}{d t}(t) \left(U^\nu - \lambda \partial^\nu \right)\delta(\bald{x} - \bald{u} t)\\
&= \int_0^T d t \frac{dE}{dt}(t)\,e^{-i k_z t + i \omega t}\left(U^\nu + i \lambda \, k^\nu \right)
\end{split}
\end{equation}
where a time derivative on $dE/dt$ as well as boundary terms have been ignored, and the source is assumed to propagate from time $t = 0$ to $t = T$.  We will discuss why we have dropped the derivative and boundary terms at the end of the results section III B.  Here we simply note that the derivative term is numerically insignificant for the energy deposition scenarios we consider, and that the boundary terms are an artifact of stopping and starting the source at a specific moment of time and obscure the physics we are trying to explore.  It is immediately clear from (\ref{momcomJ}) that the term proportional to $\lambda$ does not excite the diffusive momentum density, which is generated by the transverse part of the source, but only excites the sound modes.  The contribution to the medium excitation coming from the term proportional to $\lambda$ is important for generating Mach-like signals in the final azimuthal particle spectrum.

In what follows, we will use (\ref{momcomJ}) as our source term, treating $\lambda$ as an adjustable parameter and determining the rate of energy deposition $dE/dt$ from different theoretical models to be discussed in the next subsection.  Our purpose here is not to suggest that the form of $\lambda$ given in (\ref{lamform}) is necessarily the correct form for the QGP created in heavy-ion collisions, but rather to motivate the general form of the source, (\ref{PcomplexJ}).  The ansatz provides thus a connection between the hard, perturbative QCD physics of jet quenching and the soft, nonperturbative QCD physics of medium response.

\subsection{Energy Deposition Scenarios}\label{energy_scenarios}

In the following, we investigate three different scenarios for the energy deposition into the medium. In all three cases, we assume that the medium properties do not change as a function of space or time during the energy deposition. With this assumption, the spatiotemporal structure of the energy deposition $dE/dt$ is independent of changes in medium properties and a function of the assumed physics of the source only.  In addition, in order to study the dependence of the shockwave signal on the functional form of $dE/dt$, we adjust the medium properties for each scenario such that the {\em integrated} energy deposition $\Delta E_{tot} = \int_0^\infty dt \,dE/dt$ is the same. In contrast, if one would do calculations in a given microscopical model for parton-medium interactions and fix the medium to be e.g. at a given $T$, $\Delta E_{tot}$ would {\em not} necessarily be the same in different scenarios of energy deposition.

We adopt the procedure of normalizing to the same $\Delta E_{tot}$ nevertheless because the relevant microscopical degrees of freedom in the medium and their interaction with a hard probe are not known. In a more realistic model, with a hydrodynamically expanding medium, constraints from data on high $P_T$ observables could be utilized instead by requiring each of the energy deposition scenarios to agree with the observed suppression of high $P_T$ hadrons. In the absence of such data for a constant model, requiring the same $\Delta E_{tot}$ is a substitute for such a constraint.

In the first scenario, we assume that the source entering the medium is given by a single on-shell parton which interacts with the medium only elastically. The energy transfer into the medium is then given by the expression \cite{Thoma}
\begin{equation}
\label{E-Elastic}
\left(\frac{dE}{dt} \right)_C = \frac{\alpha_s C_2 m_D^2}{2} \ln \frac{2 \sqrt{E_p T}}{m_D}
\end{equation}
where $\alpha_s = g^2/(4\pi)$ is the strong coupling, $m_D = gT$ the Debye mass, $C_2$ the appropriate color factor for a quark (4/3) or a gluon (3) and $E_p$ the energy of the hard parton. Under the assumption that the hard parton is sufficiently energetic such that $E_p \gg \Delta E_{tot}$, the weak dependence on $E_p$ can be neglected and $\left(\frac{dE}{dt} \right)_C$ assumes a constant value if the medium temperature remains unchanged. Given this functional shape of $dE/dt$, we will refer to this scenario as {\it Flat} in the following.

\begin{figure}
\includegraphics[width = 0.96\linewidth]{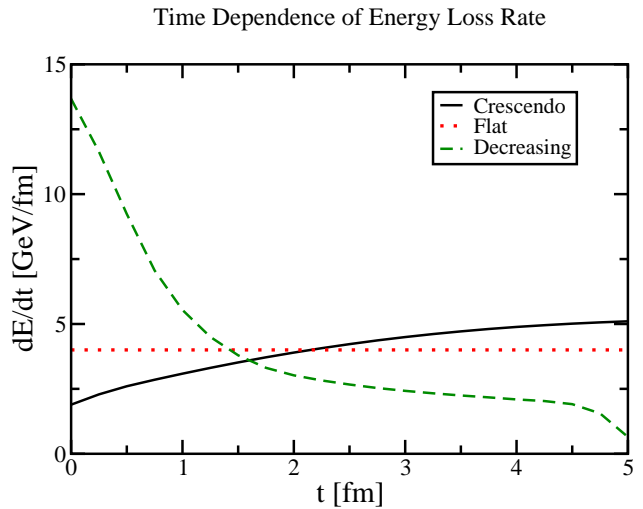}
\caption{(Color online) The time dependence of the energy deposition rate into the medium for given $\Delta E_{tot} = 20$ GeV as calculated in three different scenarios {\em Crescendo, Flat} and {\em Decreasing} for the in-medium evolution of an initial hard parton (see text).}
\label{energy_dep}
\end{figure}

In a second scenario, we still assume that the source entering the medium is a single on-shell parton with sufficient energy such that $E_p \gg \Delta E_{tot}$ is realized, but we also allow for inelastic interactions of this parton with the medium which induce radiation. As outlined in \cite{Neufeld:2009za} (see also \cite{Qin:2009uh}), the gluons radiated from the hard parent partons become themselves sources of elastic energy loss with the medium and can also be absorbed by the medium if their energy becomes $O(T)$.  

In this formalism, the effect of radiated gluons to deposit energy into the medium can be cast into the form of an evolution equation for the distribution function $f(\omega, t)$, which describes the distribution of radiated gluons in the medium at time $t$ with energy $\omega$. This evolution equation reads
\begin{equation}\label{folution}
\frac{\partial}{\partial t} f(\omega,t) - \frac{\partial}{\partial \omega} [ \epsilon(\omega) f(\omega,t)] = \frac{dI}{d\omega dt}(\omega,t)
\end{equation}
where $\epsilon(\omega)$ is the collisional energy loss rate for gluons obtained from Eq.~(\ref{E-Elastic}) as
\begin{equation}
\epsilon(\omega) = \frac{3}{2} \alpha_s m_D^2 \ln \frac{2 \sqrt{\omega T}}{m_D}
\end{equation}
and in the Armesto-Salgado-Wiedemann (ASW) formalism \cite{QuenchingWeights} the spectrum of radiated gluons is given by
\begin{equation}
\frac{dI}{d\omega dt} = - \frac{\sqrt{\hat{q}}\alpha_s C_2}{\pi} \text{Re} \frac{(1+i) \tan \left[ (1+i) \sqrt{\frac{\hat{q}}{\omega}} \frac{t}{2}\right] }{\omega^{3/2}}
\end{equation}
where
\begin{equation}
\hat{q} = 2 \alpha_s C_2 m_D^2 T \ln \frac{2\sqrt{E_p T}}{m_D}
\end{equation}
is used to adjust the strength of the inelastic interactions.

Equation (\ref{folution}) must be solved numerically. Given such a solution, the rate of energy gained by the medium from radiated gluons is \cite{Neufeld:2009za}
\begin{equation}
\begin{split}
\left(\frac{dE}{dt} \right)_R = &\int_{\omega_{min}}^{\omega_{max}} \hspace*{-3ex} d\omega \epsilon(\omega) f(\omega,t) + \int_0^{\omega_{min}}\hspace*{-3ex}d\omega \omega \frac{dI}{d\omega dt} \\&+ \omega_{min}f(\omega_{min},t) \epsilon(\omega_{min})
\end{split}
\end{equation}
where $\omega_{min} = T$ and $\omega_{max} = E_p/2$.  The total energy deposition into the medium is then given as the sum of the collisional and radiative contributions
\begin{equation}
\frac{dE}{dt}= \left(\frac{dE}{dt} \right)_C + \left(\frac{dE}{dt} \right)_R
\end{equation}
where $(dE/dt)_C$ is obtained from Eq.~(\ref{E-Elastic}).  This second scenario leads to an increase of $\frac{dE}{dt}$ in time, therefore it has been named {\it Crescendo}.

In a third scenario, we take the source entering the medium to be a highly virtual parton which subsequently evolves into a parton shower. In addition, we do not make the assumption $E_p \gg \Delta E_{tot}$ but consider finite energy kinematics for all partons. For this, we utilize the Monte Carlo (MC) code YaJEM (Yet another Jet Energy-loss Model) \cite{YAS,YAS1}. In the following, we summarize the essential parts of the computation, details can be found in \cite{YAS1} (the scenario used in this paper corresponds to the DRAG (medium-induced drag force) scenario described in \cite{YAS1}).

We model the evolution from the initial parton to a final state parton shower as a series of branching processes $a \rightarrow b+c$ where $a$ is called the parent parton and $b$ and $c$ are referred to as daughters.   In QCD, the allowed branching processes are $q \rightarrow qg$, $g \rightarrow gg$ and $g \rightarrow q \overline{q}$.  The kinematics of a branching is described in terms of the virtuality scale $Q^2$ and of the energy fraction $z$, where the energy of daughter $b$ is given by $E_b = z E_a$ and of the daughter $c$ by $E_c = (1-z) E_a$. It is convenient to introduce $t = \ln Q^2/\Lambda_{QCD}$ where $\Lambda_{QCD}$ is the scale parameter of QCD. $t$ takes a role similar to a time in the evolution equations, as it describes the evolution from some high initial virtuality $Q_0$ ($t_0$) to a lower virtuality $Q_m$ ($t_m$) at which the next branching occurs. In terms of the two variables, the differential probability $dP_a$ for a parton $a$ to branch is \cite{DGLAP1,DGLAP2}
\begin{equation}
dP_a = \sum_{b,c} \frac{\alpha_s}{2\pi} P_{a\rightarrow bc}(z) dt dz
\end{equation}
where the splitting kernels $P_{a\rightarrow bc}(z)$ read
\begin{eqnarray}
&&P_{q\rightarrow qg}(z) = 4/3 \frac{1+z^2}{1-z} \label{E-qqg}\\
&&P_{g\rightarrow gg}(z) = 3 \frac{(1-z(1-z))^2}{z(1-z)}\label{E-ggg}\\
&&P_{g\rightarrow q\overline{q}}(z) = N_F/2 (z^2 + (1-z)^2)\label{E-gqq}.
\end{eqnarray}
We do not consider electromagnetic branchings. $N_F$ counts the number of active quark flavours for given virtuality. The resulting system of equations describing the branching processes in vacuum is solved numerically using MC techniques utilizing the {\sc Pyshow} code \cite{PYSHOW}.

In order to make the link from momentum space where the shower evolution takes place to position space where the medium perturbations evolve, we assume that the average formation time of a shower parton with virtuality $Q$ is developed on the timescale $1/Q$, i.e. the average lifetime of a virtual parton with virtuality $Q_b$ coming from a parent parton with virtuality $Q_a$ is in the rest frame of the original hard collision (the local rest frame of the medium may be different by a flow boost as the medium may not be static) given by
\begin{equation}
\label{E-Lifetime}
\langle \tau_b \rangle = \frac{E_b}{Q_b^2} - \frac{E_b}{Q_a^2}.
\end{equation}
We assume that the actual formation time can then be obtained from a probability distribution
\begin{equation}
\label{E-RLifetime}
P(\tau_b) = \exp\left[- \frac{\tau_b}{\langle \tau_b \rangle}  \right]
\end{equation}
which we sample to determine the actual formation time of the fluctuation in each branching.

We assume that the medium induces an approximately constant energy loss on each propagating parton. The medium is then characterized by a drag coefficient $D$ which describes the energy loss per unit pathlength.  In the simulation, the energy (and momentum) of each propagating parton are reduced by
\begin{equation}
\label{E-Drag}
\Delta E_a = \int_{\tau_a^0}^{\tau_a^0 + \tau_a} d\zeta D
\end{equation}
For a propagating gluon the energy loss is increased by the color factor ratio 2.25. While Eq.~(\ref{E-Drag}) describes the mean energy loss, the actual energy loss due to the medium is randomized in each event.

The dynamics of energy deposition in YaJEM is rather different from the Crescendo scenario. The initial branching processes down from a highly virtual state happen very fast and lead to a pronounced initial rise in energy deposition as the number of partons undergoing elastic energy loss increases. However, the finite energy of the parton shower which is explicitly considered leads to a turnover: As partons become absorbed by the medium, the number of additionally radiated gluons is limited by kinematic constraints. As a result, the functional shape of  $\frac{dE}{dt}$ is decreasing in time, therefore the scenario is labeled {\it Decreasing} in the following.

\section{Results}\label{results}

\subsection{Input Values and Qualitative Expectations}\label{input_qual}

In the results below we consider the following situation: a source parton is created at time $t = 0$ and travels through the medium until the energy deposition ceases at the time $t = 5$ fm/c.  As discussed in section \ref{source_section}, the source parton is assumed to excite the medium through the source term (\ref{PcomplexJ}).  We continue to evolve the medium response until time $t = 7$ fm at which point the medium is assumed to hadronize into the spectrum given by (\ref{cfform}).  Performing the freeze-out $2$ fm after the source is turned off improves the validity of the linearized hydrodynamic assumption because the peak amplitude of the energy and momentum density perturbations decays in time once the source is turned off.

The underlying medium is at temperature $T_0 = 250$ MeV and has speed of sound $c_s^2 = 1/3$.  The shear viscosity to entropy density ratio, $\eta/s$, is treated as an input parameter that we vary from $\eta/s = 0.1 - 0.2$.  The range of input values for $\eta/s$ are consistent with phenomenological observations from heavy-ion collisions \cite{Romatschke:2007mq}.  The local excitation parameter, $\lambda$, mentioned in section \ref{source_section} is treated as an input parameter, which we vary between $0 - 2$ fm.  Although we treat $\lambda$ and $\eta/s$ as parameters, rigorous determination of their values must come from the underlying theory.  One may ask what sets an upper limit for $\lambda$.  From the point of view of hydrodynamics, $\lambda$ appears as the coefficient of a gradient, so in principle its value sets a minimum resolution for the hydrodynamic equations of motion.  However, in the case we are considering, we allow the medium response to decay in time, improving the validity of hydrodynamics, and in particular, linearized hydrodynamics, once the freeze-out occurs.  To quantitatively set an upper limit on $\lambda$, one must solve for the medium response for a given scenario and see how large the perturbations are.  In each of the scenarios we consider below, we find that the linearized approximation holds (to the extent that $\delta u$ appearing in (\ref{mod_dist}) remains less than 1).

The primary goal of this paper is to demonstrate that the time dependence of the energy deposition rate, $dE/dt$, which appears as a coefficient to the source (\ref{PcomplexJ}) is highly relevant for the appearance of a shockwave signal in the azimuthal spectrum (\ref{cfform}).  We will consider three different energy deposition scenarios, {\it Crescendo}, {\it Flat} and {\it Decreasing}, that have been discussed in detail in section \ref{energy_scenarios}.  In order to make the comparison consistent, in all three scenarios the total energy deposited into the medium is 20 GeV (see Figure \ref{energy_dep}).  Before examining the results of our calculations it is useful to qualitatively consider what may be expected.  From equations (\ref{mod_dist}) and (\ref{cfform}) one can see that the appearance of a shockwave signal is sensitive to the combination $\delta u(x) \, p_T/T$, where, as discussed in section \ref{azimuthal_section}, $\delta u(x)$ is the four-velocity induced by the source.  In the limit of validity of linearized hydrodynamics, $\delta u < 1$, thus, as mentioned in a previous work \cite{cst}, one expects the shockwave signal to be more pronounced for larger values of $p_T/T$.  However, it's not hard to see that this expectation remains even for nonlinear hydrodynamics.  Even if one doesn't linearize the disturbance created by the hard parton, equations (\ref{mod_dist}) and (\ref{cfform}) still predict an enhanced signal for larger $p_T$.

Experimentally, the opposite trend is seen \cite{PHENIX_2pc,STAR_2pc}: the double-peaked structure observed in the away-side distribution of di-hadron correlations is more pronounced for smaller values of $p_T$.  This observation does not mean the Mach cone shockwave is not responsible for the double-peaked structure seen experimentally, but rather exposes the limitations of hydrodynamics.  As $p_T$ increases, the driving mechanism behind the correlation structure shifts from bulk recoil to hard fragmentation.  Most events produce correlations in the low $p_T$ hydrodynamical regime, but at higher $p_T$, there's an increased bias to see the comparatively rare events in which hard pQCD is the mechanism underlying the correlation.  At $p_T$ beyond 5-6 GeV, these hard events completely dominate the visible correlation. It is this transition from soft to hard physics which governs the transition from shoulder-region to head-region in the data and is beyond the scope of our current work (see \cite{Renk:2008km} for an analysis of this transition).

\begin{figure*}
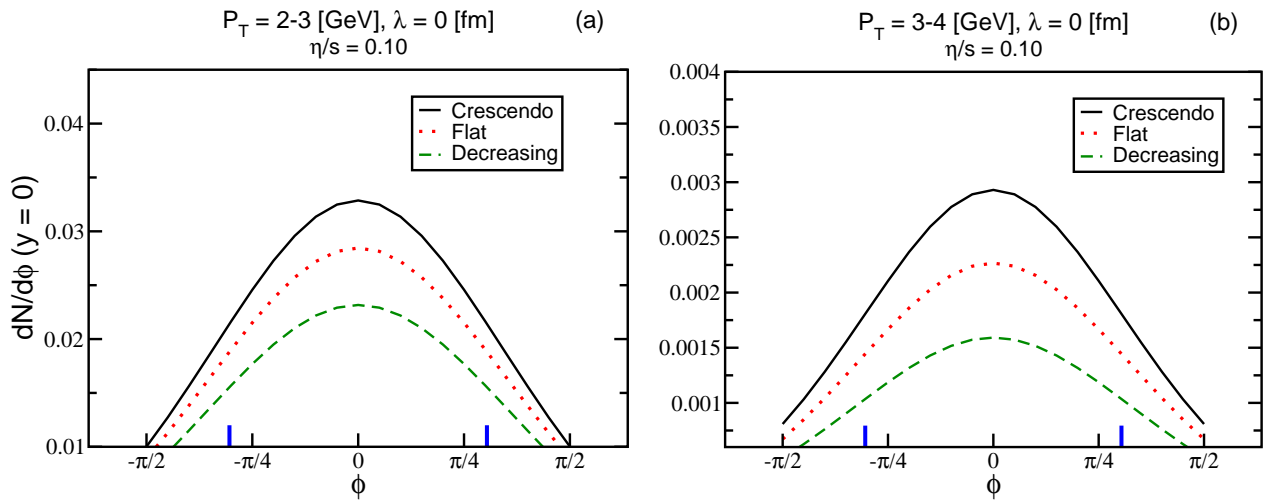

\centerline{
\includegraphics[width = 0.46\linewidth]{lam_0_pT23.eps}\hskip0.02\linewidth
\includegraphics[width = 0.45\linewidth]{lam_0_pT34.eps}
}
\caption{(Color online) The azimuthal hadron spectrum (\ref{cfform}) for the case of $\eta/s = 0.10$ and local excitation parameter, $\lambda = 0$ fm. The shapes of the spectrums for the different energy deposition scenarios are essentially the same for these parameters.  The magnitude of the signal is larger for ${\it Crescendo}$, which grows in time, than for ${\it Flat}$ or ${\it Decreasing}$, even though the same total energy is deposited in each case.  The conical structure generated by the source (\ref{PcomplexJ}) when $\lambda = 0$ fm is not strong enough to overcome the diffusive wake in the final spectrums, where one sees a single peak at $\phi = 0$, which defines the direction of source propagation.  The larger (blue) tick marks on the $\phi$ axis indicate where one would naively expect conical peaks to appear for the speed of sound used here, $\phi = \arccos c_s$.}
\label{lam_0_pT}
\end{figure*}

\begin{figure*}
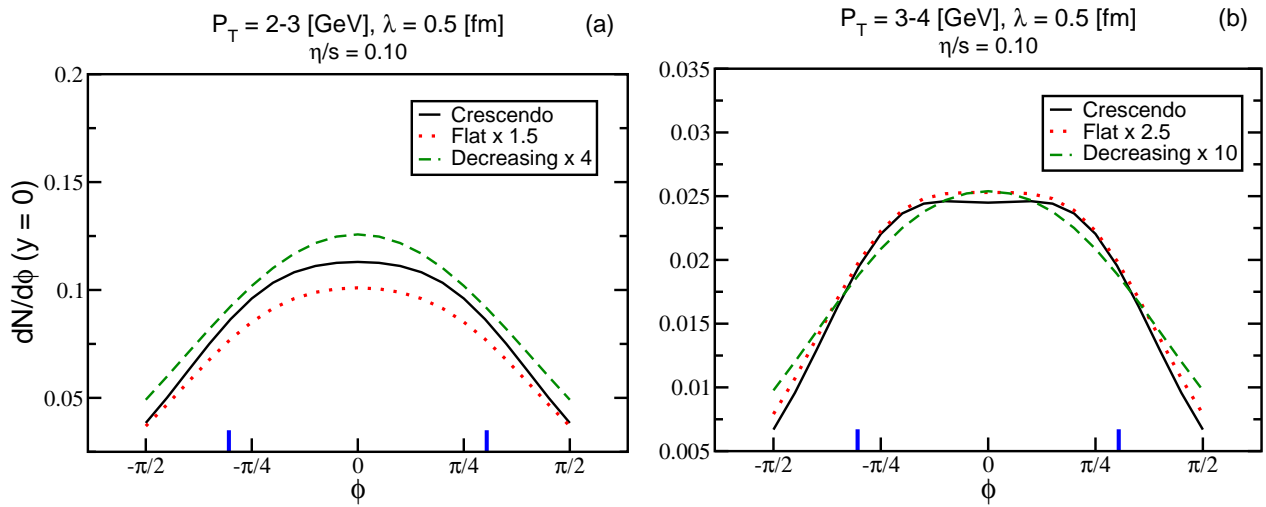

\centerline{
\includegraphics[width = 0.46\linewidth]{lam_050_pT23.eps}\hskip0.02\linewidth
\includegraphics[width = 0.45\linewidth]{lam_050_pT34.eps}
}
\caption{(Color online) The same as in Figure \ref{lam_0_pT}, but now $\lambda = 0.5$ fm. The spectrum is noticeably flatter than seen in the $\lambda = 0$ fm case, especially for the $p_T = 3 - 4$ GeV plot, where a double-peak begins to emerge in the {\it Crescendo} curve.  The difference in the shapes of the spectrums generated by the different energy deposition scenarios is most noticeable when comparing the {\it Crescendo} and {\it Decreasing} scenarios.  The energy deposition which grows in time generates a stronger conical signal.  The larger (blue) tick marks on the $\phi$ axis indicate where one would naively expect conical peaks to appear for the speed of sound we have used.}
\label{lam_050_pT}
\end{figure*}

Generally speaking, one would expect that larger amplitudes of the induced energy and momentum density perturbations would be more likely to generate a shockwave signal in the azimuthal spectrum than smaller ones.  In terms of the time dependence of the energy deposition rate, $dE/dt$, the naive expectation is that an energy deposition rate that grows in time is more favorable for the appearance of a shockwave signal than one which decreases or is flat.  To make the argument more concrete, consider a source that deposits some amount of energy and momentum at an instant in time and is then turned off.  The time dependence of the medium perturbation due to such a source can be tracked by going back to (\ref{eps} - \ref{gt}) and using the reverse Fourier transform (\ref{generalrule}).  Performing the $\omega$ integration using contour methods and leaving the $\bald{k}$ integration undone shows that, to first order in shear viscosity, as a function of time
\eqf{
\delta \epsilon, \, g_L \sim \exp \left[-\frac{\Gamma_s\,k^2 \, t}{2}\right]
}
and
\eqf{
g_T \sim \exp \left[-\frac{3 \, \Gamma_s \,k'^2 \, t}{4}\right].
}
\begin{figure*}
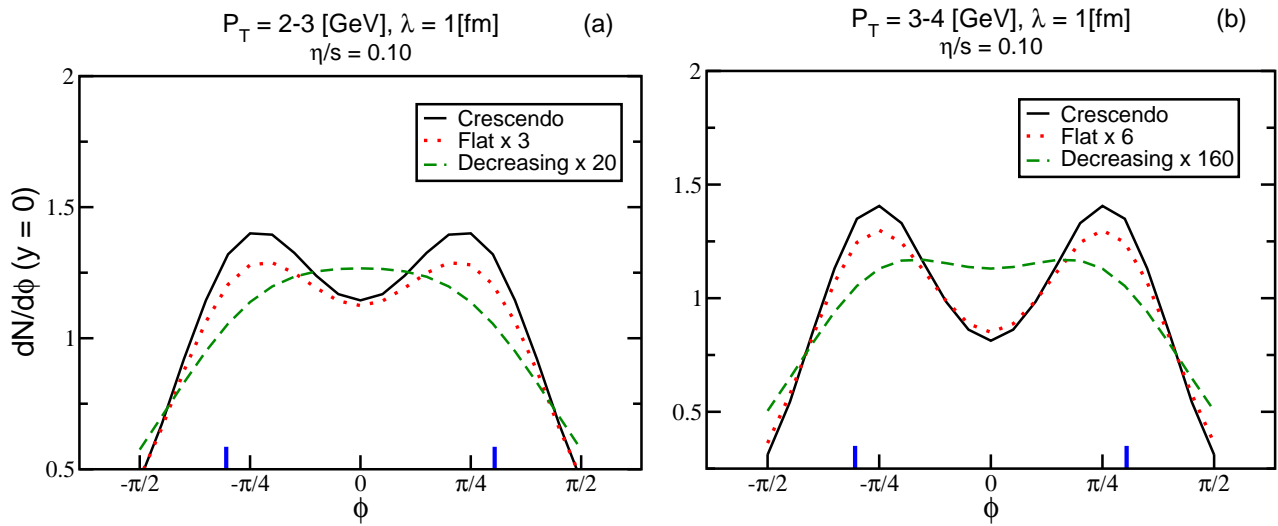

\centerline{
\includegraphics[width = 0.47\linewidth]{lam_1_pT23.eps}\hskip0.02\linewidth
\includegraphics[width = 0.45\linewidth]{lam_1_pT34.eps}
}
\caption{(Color online) The same as in Figures \ref{lam_0_pT} and \ref{lam_050_pT}, but now $\lambda = 1.0$ fm.  The double-peaked structure is significantly enhanced, particularly in the {\it Crescendo} and  {\it Flat} spectrums.   The difference in the shapes of the spectrums generated by the different energy deposition scenarios is obvious.  The larger (blue) tick marks on the $\phi$ axis indicate where one would naively expect conical peaks to appear for the speed of sound used here.}
\label{lam_1_pT}
\end{figure*}

\begin{figure*}
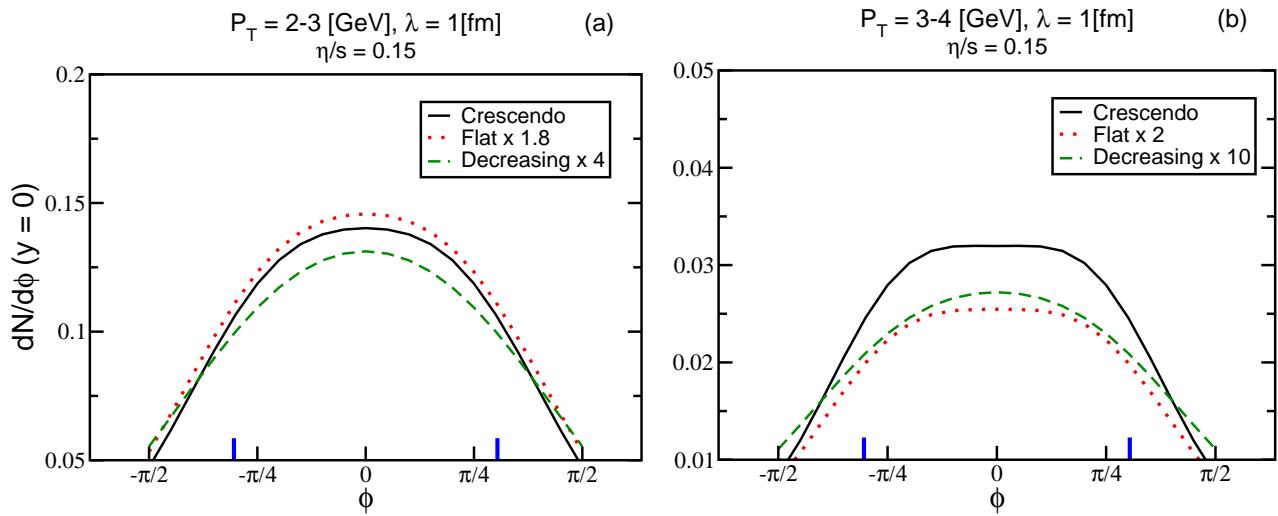

\centerline{
\includegraphics[width = 0.47\linewidth]{lam_1_pT23_II.eps}\hskip0.02\linewidth
\includegraphics[width = 0.45\linewidth]{lam_1_pT34_II.eps}
}
\caption{(Color online) The azimuthal hadron spectrum (\ref{cfform}) for the case of $\eta/s = 0.15$ and $\lambda = 1.0$ fm.  The effect of viscosity to smear out any double-peaked structure is obvious when comparing to the results in Figure \ref{lam_1_pT}, where we considered  $\eta/s = 0.10$ and $\lambda = 1.0$ fm.  The larger (blue) tick marks on the $\phi$ axis indicate where one would naively expect conical peaks to appear for the speed of sound used here.}
\label{lam_1_pT_II}
\end{figure*}

The medium excitation decays exponentially as a function of time, meaning that the azimuthal spectrum (\ref{cfform}) more reflects the strength of energy deposition at later times than earlier times.  However, the situation is complicated by the fact that the diffusive contribution, $g_T$, decays more quickly in time than the sound contribution, $\delta \epsilon$ and $g_L$.  It was mentioned in section \ref{source_section} that the appearance of a shockwave signal in the azimuthal spectrum will come from $\delta \epsilon$ and $g_L$, whereas $g_T$ tends to fill up any double-peak structure.  Even though the amplitude of the medium excitation decays exponentially in time for both the sound and diffusion modes, the ratio of the sound/diffusive contribution actually grows in time.  From these considerations, it is not immediately clear whether an energy deposition rate which grows in time or decreases in time is more favorable for generating a shockwave signal.  In the next subsection we present the results of our calculations.

\subsection{Numerical Results}\label{num_results}

\begin{figure*}
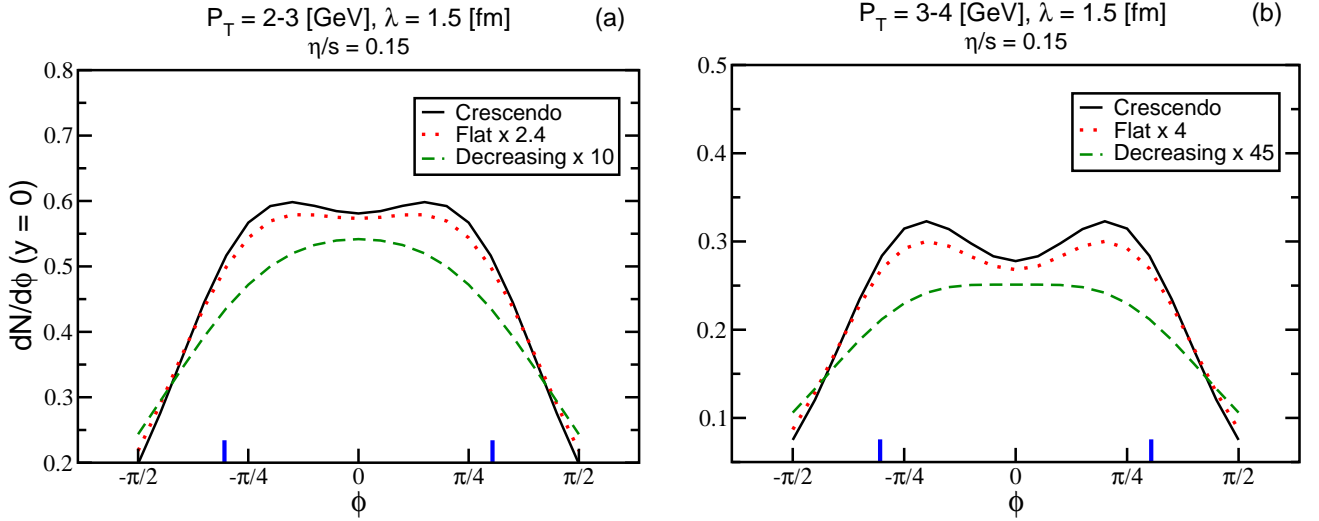

\centerline{
\includegraphics[width = 0.47\linewidth]{lam_150_pT23.eps}\hskip0.04\linewidth
\includegraphics[width = 0.45\linewidth]{lam_150_pT34.eps}
}
\caption{(Color online) The azimuthal hadron spectrum for $\eta/s = 0.15$ and $\lambda = 1.5$ fm.  Increasing $\lambda$ restores the shockwave signal in the {\it Flat} and {\it Crescendo} energy deposition scenarios, (compare to Figure \ref{lam_1_pT_II} where $\lambda = 1.0$).  However, consistent with the above results, the {\it Decreasing} energy deposition remains mostly flat.  The larger (blue) tick marks on the $\phi$ axis indicate where one would naively expect conical peaks to appear for the speed of sound used here.}
\label{lam_150_pT}
\end{figure*}

\begin{figure*}
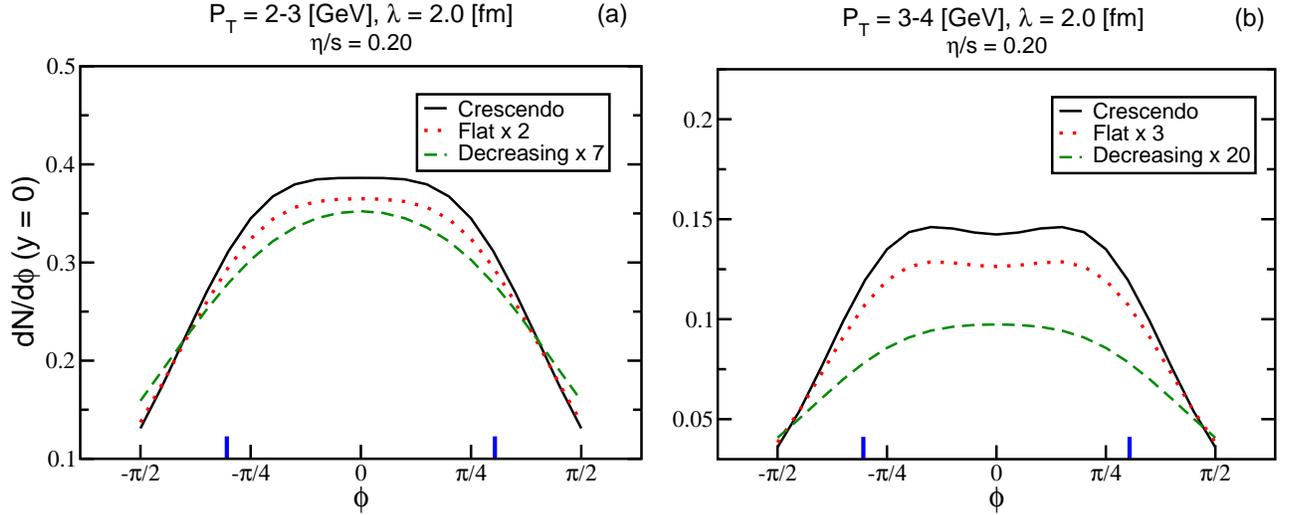

\centerline{
\includegraphics[width = 0.47\linewidth]{lam_2_pT23.eps}\hskip0.02\linewidth
\includegraphics[width = 0.45\linewidth]{lam_2_pT34.eps}
}
\caption{(Color online) The resulting spectrums for $\eta/s = 0.20$ and $\lambda = 2.0$ fm. Again, the result of increasing the viscosity is to smear out the double-peaked structure, even though we have increased $\lambda$ by the same fraction as presented in Figure \ref{lam_150_pT}.  Recall that the larger (blue) tick marks on the $\phi$ axis indicate where one would naively expect conical peaks to appear for the speed of sound used here.}
\label{lam_2_pT}
\end{figure*}

\begin{figure*}
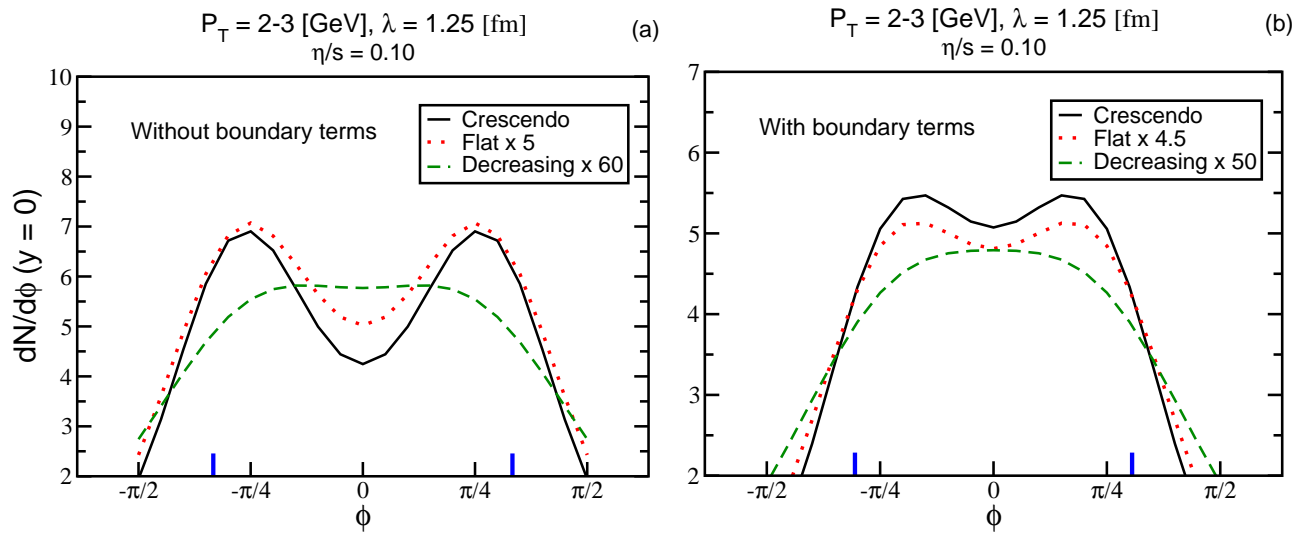

\centerline{
\includegraphics[width = 0.48\linewidth]{nobound.eps}\hskip0.02\linewidth
\includegraphics[width = 0.45\linewidth]{bound.eps}
}
\caption{(Color online) Results for $p_T = 2 - 3$ GeV, $\eta/s = 0.10$ and $\lambda = 1.25$ fm.  The left panel shows the result without including the boundary and derivative terms (as discussed in the text), whereas the right panel contains those contributions.  In the right panel, one sees that the difference in the shapes of the curves are less significant, and also narrower.  This results from inward flow created by the absorption of the source at an instant in time, and is an artifact of an infinite energy (for the case of the {\it Crescendo} and {\it Flat} scenarios), static medium assumption.  We have not included the boundary terms in the above results, as they obscure the physics we are trying to study.  The larger (blue) tick marks on the $\phi$ axis indicate where one would naively expect conical peaks to appear for the speed of sound used here.}
\label{boundbound}
\end{figure*}

In this section we present results for the azimuthal hadron spectrum (\ref{cfform}) obtained using the medium parameters and energy deposition scenarios discussed above.  In all results we will show the spectrum for the three energy deposition scenarios, {\it Crescendo}, {\it Flat} and {\it Decreasing}, and for the bins $p_T = 2 - 3$ and $p_T = 3 - 4$ GeV. 
We now present results for the azimuthal hadron spectrum (\ref{cfform}) obtained using the medium parameters and energy deposition scenarios discussed above.  In all results we will show the spectrum for the three energy deposition scenarios, {\it Crescendo}, {\it Flat} and {\it Decreasing}, and for the bins $p_T = 2 - 3$ and $p_T = 3 - 4$ GeV.  Results for shear viscosity to entropy density ratio $\eta/s = 0.10$ and local excitation parameter, $\lambda = 0$ fm ($\lambda$ is discussed in section \ref{source_section}) are shown in Figure \ref{lam_0_pT}.  Recall that the direction of the source propagation determines the direction of $\phi = 0$ in our plots.  The results show that the shape of the spectrum is roughly the same for all three energy deposition scenarios, however, the magnitude of the signal is larger for ${\it Crescendo}$, which grows in time, than for ${\it Flat}$ or ${\it Decreasing}$.  This change in magnitude between the different energy deposition scenarios reflects the viscous nature of the medium.  Energy which is deposited at earlier times (such as in the ${\it Decreasing}$ scenario) has more time to equilibrate with the background medium before freeze-out.  What is noticeably missing in Figure \ref{lam_0_pT} is the appearance of a double-peaked structure, or shockwave signal.  Apparently the conical structure generated by the source (\ref{PcomplexJ}) when $\lambda = 0$ fm is not strong enough to overcome the diffusive wake in the final spectrum.

Next we consider results for $\eta/s = 0.10$ and $\lambda = 0.5$ fm which are shown in Figure \ref{lam_050_pT}.  The spectrum is noticeably flatter than seen in the $\lambda = 0$ fm case, especially for the $p_T = 3 - 4$ GeV plot, where a double-peak begins to emerge in the {\it Crescendo} curve.  Notice that not only are the magnitudes of the curves for the different energy deposition scenarios different, but also the shapes.  The difference is most noticeable when comparing the {\it Crescendo} and {\it Decreasing} scenarios.  The shape of the {\it Decreasing} spectrum is very similar to the case of $\lambda = 0$ fm, however the shape of the {\it Crescendo} spectrum is much flatter, even showing a slight dip at $\phi = 0$.  The energy deposition which grows in time appears to generate a cleaner signal of the underlying conical structure induced in the medium.

We continue by considering results for $\eta/s = 0.10$ and $\lambda = 1.0$ fm presented in Figure \ref{lam_1_pT}.   Here the double-peaked structure becomes quite pronounced for the {\it Flat}  and {\it Crescendo} curves, however the {\it Decreasing} result remains mostly flat, with a slight dip at $\phi = 0$ in the $p_T = 3-4$ GeV range.  The  {\it Flat}  and {\it Crescendo} spectrums have a similar shape, however the {\it Crescendo} result shows a moderately more pronounced double-peak.  The trend is emerging that for fixed values of $\eta/s$ and increasing values of $\lambda$, an energy deposition scenario which increases in time generates a more pronounced shockwave signal in the final azimuthal spectrum.

The results for $\eta/s = 0.15$ and $\lambda = 1.0$ fm are shown in Figure \ref{lam_1_pT_II}.  One can see that a 50 percent increase in the shear viscosity has a significant effect on the final azimuthal spectrum by comparing with Figure \ref{lam_1_pT}.  The double-peaked shockwave signature has been smeared out by the viscous effects.  Continuing with $\eta/s = 0.15$ and increasing $\lambda$ to 1.5 fm restores the shockwave signal in the {\it Flat} and {\it Crescendo} energy deposition scenarios, as is evident from Figure \ref{lam_150_pT}.  However, consistent with the above results, the {\it Decreasing} energy deposition remains mostly flat.  Finally, we present results for $\eta/s = 0.20$ and $\lambda = 2.0$ fm in Figure \ref{lam_2_pT}.  Again, the result of increasing the viscosity is to smear out the double-peaked structure, even though we have increased $\lambda$ by the same fraction.

We here discuss the form of the source, (\ref{momcomJ}), which we have employed for the results presented above.  As mentioned briefly in section II C, we do not include a time derivative on $dE/dt$ or boundary terms which arise from the source being turned on and off at $t = 0$ and $t = T$.  As noted above, we have explicitly checked that the derivative term is numerically insignificant for the energy deposition scenarios we consider and can safely be ignored.  However, the boundary terms, in particular, the term resulting from turning the source off at time $t = T$, is not numerically insignificant.  In fact, the effect of the boundary term at $t = T$ is to create a strong inward flow as the source is absorbed by the medium.  This inward flow tends to destroy the conical Mach cone signal, and reduce the differences in the three scenarios.

This feature is demonstrated in Figure \ref{boundbound} where we show the azimuthal spectrum for the bin $p_T = 2 - 3$ GeV, $\eta/s = 0.10$ and $\lambda = 1.25$ fm.  The left panel shows the result without including the boundary and derivative terms, whereas the right panel contains those contributions (these are dominated by the $t = T$ boundary piece).  The left panel is characteristic of the plots shown above, in that the {\it Crescendo} curve has the most pronounced double peak structure, whereas the {\it Flat} and {\it Decreasing} curves are less pronounced.  In the right panel, however, one sees that the difference in the shapes of the curves are less significant, and also narrower.  This results directly from inward flow created by the absorption of the source at an instant in time.

The absorption at $t = T$ is an artifact of our procedure to stop the simulation, and tends to obscure the physics we are trying to study.  In a real physical situation, the source is not simply absorbed at some fixed instant, but may escape the medium or run out of energy, etc.  With an infinite energy (for the case of the {\it Crescendo} and {\it Flat} scenarios), static medium assumption, the problem of stopping the energy deposition is not well defined (it would go on forever), so we have to make a choice as to what we want to show as a freeze-out distribution.  We believe we can get closer to what we're interested in by dropping the boundary terms - which we have done in the above.  However, the complications just described illustrate the need for a more realistic description of both the medium evolution and the source term.

\begin{figure}
\centerline{
\includegraphics[width = 0.96\linewidth]{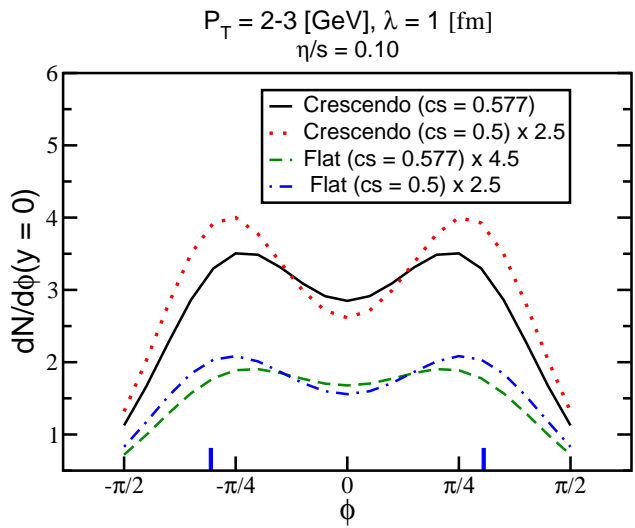}
}
\caption{(Color online) Results for $p_T = 2 - 3$ GeV, $\eta/s = 0.10$ and $\lambda = 1.0$ fm with two different speeds of sound.  As discussed in the text, one would anticipate the spectrum becomes more pronounced when lowering the speed of sound and that the peaks would appear at larger opening angles.  These two features are indeed observed.  However, notice that the peak angles do not correspond to a naive prediction based on purely geometric arguments (the blue ticks in the Figure would naively correspond to  $c_s = 0.577$) even for the simple scenario considered here.}
\label{speedandsoundbaby}
\end{figure}

We conclude this section by briefly considering how our results depend on the specific choice of speed of sound $c_s^2 = 1/3$ and temperature $T_0 = 250$ MeV we have used here.  Both the speed of sound and the temperature appear explicitly in the medium distribution function (see (\ref{pertu}) and (\ref{mod_dist})) used for freeze-out.  Changing the temperature in the medium distribution function (\ref{mod_dist}) will have an effect on the strength of the freeze-out signal.  Specifically, lowering the temperature will tend to generate a stronger signal.  Likewise, lowering the speed of sound in the medium distribution function will also tend to create a stronger freeze-out signal (even more so than the temperature, since the speed of sound appears only with the momentum flow).

The temperature also enters the sound attenuation length, $\Gamma_s = \frac{4 \eta }{3 s T}$, which appears in the equations of motion for linearized hydrodynamics (\ref{eps} - \ref{gt}).  Changing the temperature in $\Gamma_s$ is effectively like changing $\eta/s$, which has been analyzed in the results above, thus we do not consider that aspect further here.  However, the speed of sound appears in a non-trivial way in the equations of motion.  As is well understood, the speed of sound governs the angle of propagation of Mach cone shock waves generated by a projectile.  We present in Figure \ref{speedandsoundbaby} the effects of changing the speed of sound on the the azimuthal hadron spectrum for the {\it Crescendo} and {\it Flat} energy deposition scenarios and for the bin $p_T = 2 - 3$ GeV.  The results are for $\eta/s = 0.10$, $\lambda = 1$ fm and speed of sound $c_s = 0.577$ and $c_s = 0.5$.  As anticipated, the spectrum becomes more pronounced when lowering the speed of sound, and the peaks also appear at larger opening angles.  However, notice that the peak angles do not correspond to a naive prediction based on purely geometric arguments even for the simple scenario considered here (the blue ticks in the Figure would naively correspond to  $c_s = 0.577$). One thus needs to be very careful not to interpret the experimentally measured opening angle of the correlation geometrically as directly related to the speed of sound --- in a realistic medium, the combination of trigger bias with a longitudinal, transverse and elliptic flow field will have an even more significant influence on the angular structure than in our simplified medium study.

\section{Conclusions}

We have conducted a systematic study of the shockwave excitation for different models of the spatio-temporal structure of energy deposition into the medium within a linearized hydrodynamical framework in a constant medium under different assumptions with regard to medium properties and the interaction of the source with the medium. The results exhibit a few generic trends:

\begin{itemize}
\item Only a strong gradient term ($\lambda > 0$) in the source (\ref{PcomplexJ}) leads to an observable double-hump structure.  The observation that a gradient term is necessary to excite an observable double-peak has been made early on \cite{solana}, however, our implementation of the source (\ref{PcomplexJ}) provides a way to quantify how strong the gradient term must be.
\item For fixed strength of the gradient term, viscous effects (i.e. larger values of $\eta/s$) weaken the double-hump structure. This has a natural explanation in terms of entropy generation dissipating the shockwave, but even for relatively small $\eta/s$ the effect appears rather pronounced.
\item Consistently for all assumptions about the structure of the source term and the medium shear viscosity, an energy deposition $dE/dt$ which increases as a function of time leads to more pronounced shockwave-like correlations than $dE/dt$ decreasing in time --- both in the absolute strength of the correlation as well as in the shape. This effect is not small --- the correlation strength can be reduced more than an order of magnitude in the 'Decreasing' as compared to the 'Crescendo' scenario, although the precise factor depends on the medium properties.
\end{itemize}

Applied to the shockwave interpretation of measured correlations in heavy-ion collisions, these findings imply that the measured signal strongly depend on medium properties, the local structure of the source term and the spatio-temporal pattern of energy deposition. Thus, if one is able to model the evolution of an initial hard parton and its interaction with the medium with sufficient precision, one can determine medium parameters like $\eta/s$ from the measured correlations. On the other hand, if one can extract the medium properties with other methods, the correlations then place tight constraints on the dynamics of hard parton evolution in the medium. Which road will be taken first still remains to be seen.

\begin{acknowledgments}
This work was supported by an Academy Research Fellowship from the Finnish Academy and from Academy Project 115262, and also by 
the US Department of Energy, Office of Science, under Contract No. DE-AC52-06NA25396.
\end{acknowledgments}


\begin{thebibliography}{99}

\bibitem{Jet1}
  M.~Gyulassy and X.~N.~Wang,
  Nucl.\ Phys.\ B {\bf 420}, (1994) 583.

\bibitem{Jet2}
  R.~Baier, Y.~L.~Dokshitzer, A.~H.~Mueller, S.~Peigne and D.~Schiff,
  Nucl.\ Phys.\ B {\bf 484}, (1997) 265.

\bibitem{Jet3}
  B.~G.~Zakharov,
  JETP Lett.\  {\bf 65}, (1997) 615.

\bibitem{Jet4}
  U.~A.~Wiedemann,
  Nucl.\ Phys.\ B {\bf 588}, (2000) 303.

\bibitem{Jet5}
  M.~Gyulassy, P.~Levai and I.~Vitev,
  Nucl.\ Phys.\ B {\bf 594}, (2001) 371.

\bibitem{Jet6}
  X.~N.~Wang and X.~F.~Guo,
  Nucl.\ Phys.\ A {\bf 696}, (2001) 788.

\bibitem{PHENIX_R_AA}
  M.~Shimomura  [PHENIX Collaboration],
  nucl-ex/0510023.

\bibitem{PHENIX_2pc}
 S.~S.~Adler {\it et al.}  [PHENIX Collaboration],
  Phys.\ Rev.\ Lett.\  {\bf 97} (2006) 052301.

\bibitem{STAR_2pc}
  J.~Adams {\it et al.}  [STAR Collaboration],
  Phys.\ Rev.\ Lett.\  {\bf 95}, 152301 (2005); J.~G.~Ulery  [STAR Collaboration],
  Nucl.\ Phys.\ A {\bf 774} (2006) 581.

\bibitem{solana}
  J.~Casalderrey-Solana, E.~V.~Shuryak and D.~Teaney,
  J.\ Phys.\ Conf.\ Ser.\  {\bf 27}, 22 (2005);

\bibitem{Cones1}
  T.~Renk and J.~Ruppert,
  Phys.\ Rev.\  C {\bf 73} (2006) 011901.

\bibitem{Cones2}
  T.~Renk and J.~Ruppert,
  Phys.\ Lett.\  B {\bf 646} (2007) 19.

\bibitem{Cones3}
 T.~Renk and J.~Ruppert,
  Phys.\ Rev.\  C {\bf 76} (2007) 014908.

\bibitem{Neufeld:2008fi}
  R.~B.~Neufeld, B.~M\"uller and J.~Ruppert,
  Phys.\ Rev.\  C {\bf 78}, 041901 (2008);
  B.~Muller and R.~B.~Neufeld,
  J.\ Phys.\ G {\bf 35}, 104108 (2008).

\bibitem{Neufeld:2008hs}
  R.~B.~Neufeld,
  Phys.\ Rev.\  D {\bf 78}, 085015 (2008).

\bibitem{Neufeld:2008dx}
  R.~B.~Neufeld,
  Phys.\ Rev.\  C {\bf 79}, 054909 (2009).

\bibitem{Betz:2008wy}
  B.~Betz, M.~Gyulassy, J.~Noronha and G.~Torrieri,
  Phys.\ Lett.\  B {\bf 675}, 340 (2009).
  
\bibitem{Noronha:2008tg}
  J.~Noronha and M.~Gyulassy,
  0806.4374 [hep-ph].
  
\bibitem{Neufeld:2009za}
  R.~B.~Neufeld and B.~Muller,
  Nucl.\ Phys.\  A {\bf 830}, 789C (2009);
  R.~B.~Neufeld and B.~Muller,
  Phys.\ Rev.\ Lett.\  {\bf 103}, 042301 (2009).

\bibitem{Qin:2009uh}
  G.~Y.~Qin, A.~Majumder, H.~Song and U.~Heinz,
  Phys.\ Rev.\ Lett.\  {\bf 103}, 152303 (2009)
  [arXiv:0903.2255 [nucl-th]].

\bibitem{Bearden:2003fw}
  I.~G.~Bearden {\it et al.}  [BRAHMS Collaboration],
  Phys.\ Rev.\ Lett.\  {\bf 90}, 102301 (2003).

\bibitem{Heinz:2009xj}
  U.~W.~Heinz,
  arXiv:0901.4355 [nucl-th].


\bibitem{Cooper:1974mv}
  F.~Cooper and G.~Frye,
  Phys.\ Rev.\  D {\bf 10}, 186 (1974).


\bibitem{Betz:2008ka}
  B.~Betz, J.~Noronha, G.~Torrieri, M.~Gyulassy, I.~Mishustin and D.~H.~Rischke,
  Phys.\ Rev.\  C {\bf 79}, 034902 (2009).

\bibitem{Chesler:2007sv}
  P.~M.~Chesler and L.~G.~Yaffe,
  Phys.\ Rev.\  D {\bf 78}, 045013 (2008).

\bibitem{Thoma}
  M.~H.~Thoma, Phys.\ Lett.\ B {\bf 273}, 128 (1991).

\bibitem{QuenchingWeights}
  C.~A.~Salgado and U.~A.~Wiedemann,
  Phys.\ Rev.\ D {\bf 68}, (2003) 014008.

\bibitem{YAS}
T.~Renk,
  Phys.\ Rev.\  C {\bf 78} (2008) 034908.

\bibitem{YAS1}
  T.~Renk,
  Phys.\ Rev.\  C {\bf 79} (2009) 054906.

\bibitem{DGLAP1}
  V.~N.~Gribov and L.~N.~Lipatov, Sov.\ J.\ Nucl.\ Phys.\ {\bf 15} (1972) 438, {\em ibid.} {\bf 75};
Yu.~L.~Dokshitzer, Sov.\ J.\ Phys.\ JETP {\bf 46} (1977) 641.

\bibitem{DGLAP2}
  G.~Altarelli and G.~Parisi, Nucl.\ Phys.\ B {\bf 126} (1977) 298.

\bibitem{PYSHOW}
  M.~Bengtsson and T.~Sj\"{o}strand, Phys.\ Lett.\ B {\bf 185} (1987) 435; Nucl. Phys.
B {\bf 289} (1987) 810; E.~Norrbin and T.~Sj\"{o}strand, Nucl.\ Phys.\ B {\bf 603} (2001) 297.

\bibitem{Romatschke:2007mq}
  P.~Romatschke and U.~Romatschke,
  Phys.\ Rev.\ Lett.\  {\bf 99}, 172301 (2007).

\bibitem{cst}
  J.~Casalderrey-Solana, E.~V.~Shuryak and D.~Teaney,
  hep-ph/0602183.

\bibitem{Renk:2008km}
  T.~Renk,
  Phys.\ Rev.\  C {\bf 78}, 014903 (2008)
  [arXiv:0804.1204 [hep-ph]].


\end{thebibliography}
\end{document}